\documentclass[10pt]{article}

\usepackage{authblk}

\title{Human-Like Trajectories Generation via Receding Horizon Tracking Applied to the TickTacking Interface}

\author[1]{Daniele Masti}
\author[2]{Stefano Menchetti}
\author[3]{\c{C}a\u{g}r{\i} Erdem}
\author[2]{Giorgio Gnecco}
\author[3]{Davide Rocchesso}

\affil[1]{Gran Sasso Science Institute, Viale F. Crispi 7, 67100 L’Aquila, Italy\\
\texttt{daniele.masti@gssi.it}}

\affil[2]{IMT School for Advanced Studies Lucca, Piazza San Francesco 19, 55100 Lucca, Italy\\
\texttt{\{stefano.menchetti, giorgio.gnecco\}@imtlucca.it}}

\affil[3]{University of Milan, Via Celoria 18, 20133 Milan, Italy\\
\texttt{\{cagri.erdem, davide.rocchesso\}@unimi.it}}

\date{} 

\newcommand{\titlefootnote}[1]{%
  \begingroup
  \renewcommand\thefootnote{}\footnotetext{#1}%
  \addtocounter{footnote}{-1}%
  \endgroup
}

\usepackage{graphicx, scalerel}
\usepackage{url}
\usepackage{amsmath, amsfonts, amssymb}
\usepackage{float}
\usepackage{placeins}
\usepackage{glossaries}
\usepackage{array}
\usepackage{cite}
\usepackage{subfigure}
\usepackage{booktabs}
\usepackage{musixtex}
\usepackage{hyperref}
\usepackage[most]{tcolorbox}

\newacronym{rh}{RH}{Receding Horizon}
\newacronym{mpc}{MPC}{Model Predictive Control}
\newacronym{kde}{KDE}{Kernel Density Estimate}
\newacronym{rmse}{RMSE}{Root Mean Squared Error}
\newacronym{mse}{MSE}{Mean Squared Error}
\newacronym{mae}{MAE}{Mean Absolute Error}
\newacronym{pdf}{PDF}{Probability Density Function}
\newacronym{lp}{LP}{Linear Programming}
\newacronym{qp}{QP}{Quadratic Programming}
\newacronym{miqp}{MIQP}{Mixed Integer Quadratic Programming}
\newacronym{cdf}{CDF}{Cumulative Density Function}
\newacronym{hci}{HCI}{Human-Computer Interface}
\newacronym{hcis}{HCIs}{Human-Computer Interfaces}
\newacronym{ioi}{IoI}{Inter-Onset Interval}
\newacronym{lqr}{LQR}{Linear Quadratic Regulator}

\begin{document}

\maketitle
\titlefootnote{*This work was partially supported by: the PRIN 2022 project ``MAHATMA'' (CUP: D53D23008790006); by the PRIN PNRR 2022 project ``MOTUS'' (CUP: D53D23017470001); and by Italian Ministry of University and Research - National Innovation Ecosystem grant ECS00000041 - VITALITY (CUP: D13C21000430001). All projects and grants have been funded by the European Union -- Next Generation EU program.}

\thispagestyle{empty}
\pagestyle{empty}

\begin{abstract}

TickTacking is a rhythm-based interface that allows users to control a pointer in a two-dimensional space through dual-button tapping. This paper investigates the generation of human-like trajectories using a receding horizon approach applied to the TickTacking interface in a target-tracking task. By analyzing user-generated trajectories, we identify key human behavioral features and incorporate them in a controller that mimics these behaviors. The performance of this human-inspired controller is evaluated against a baseline optimal-control-based agent, demonstrating the importance of specific control features for achieving human-like interaction. These findings contribute to the broader goal of developing rhythm-based human-machine interfaces by offering design insights that enhance user performance, improve intuitiveness, and reduce interaction frustration.

\end{abstract}
\noindent\rule{8.4cm}{1pt}\\
This is the authors' version of a paper accepted at \textit{2025 IEEE International Conference on Systems, Man and Cybernetics (SMC)}. It is posted here for your personal use, not for redistribution.\\

You may use the following bibtex entry:
\begin{tcolorbox}
@INPROCEEDINGS\{11342906, \\
  author=\{Masti, Daniele and Menchetti, Stefano and Erdem, Çağrı\\
  and Gnecco, Giorgio and Rocchesso, Davide\}, \\
  booktitle=\{2025 IEEE International Conference on Systems, Man, \\
  and Cybernetics (SMC)\}, \\
  title=\{Human-Like Trajectories Generation via Receding Horizon Tracking Applied to the TickTacking Interface\}, \\
  year=\{2025\}, \\
  volume=\{\}, \\
  number=\{\}, \\
  pages=\{1476-1481\}, \\
  doi=\{10.1109/SMC58881.2025.11342906\} \\
  \}
\}
\end{tcolorbox}
\noindent\rule{8.4cm}{1pt}
\section{INTRODUCTION}

    \gls{hcis} rely on multiple input and output modalities to facilitate interaction between users and computers~\cite{10046656}.
    In most cases, they convey information through visual and auditory signals, while receiving input via keyboards, buttons, and other pressure-based methods. However, interaction is not limited to these conventional inputs, as variations in timing, rhythm, and input sequences can also serve as valuable sources of information.

    A notable example of rhythm-based HCI is the \emph{TickTacking} interface, introduced in~\cite{rocchesso2023ticktacking} and further detailed in~\cite{rocchesso2024spacetime}. This interface falls within the broader research area of rhythmic interaction in human-computer interaction, where rhythm serves as a communication channel between users and computers~\cite{erkut201317}. 
    In~\cite{rocchesso2024spacetime}, the TickTacking interface was tested in a gamified scenario where users tracked an object on the screen using a pointer controlled by rhythmic tapping on two buttons. Although this study highlighted high user engagement, the effectiveness of the interface in conveying precise commands has not been systematically evaluated. This work addresses that gap by comparing an optimal-control-based agent designed for efficiency with an artificial agent that mimics human behavior when using the TickTacking interface.

    Focusing on designing a controller that exhibits human-like behavior rather than directly analyzing human trajectories serves several key purposes. First, it provides insight into the trade-offs, constraints, and objectives that shape human interaction with the TickTacking interface. These insights can inform interface improvements and the development of user guidance systems that improve performance and intuitiveness, and reduce frustration by providing visual or auditory feedback. In addition, such guidance systems can be integrated into a cyclic learning process, where they help users refine their control strategies, generating new data that are leveraged to improve the guidance systems. Further, a human-inspired controller enables the prediction of user outcomes in novel but related tasks on the same interface. Ultimately, this research contributes to the broader goal of identifying fundamental principles for designing general-purpose rhythmic HCIs, building on insights from previous works such as~\cite{di2013stochastic,lazcano2021mpc,cecchin2024real,luo2022human}.
    
    This paper is organized as follows:~\S~\ref{sec:ticktacking} provides an overview of the TickTacking interface, explaining its functionality and how human-generated trajectories are collected. We analyze these trajectories in two scenarios: one in which users freely navigate and the other in which they perform a target-tracking task, extracting key behavioral features from both. \S~\ref{sec:receding_horizon} describes the receding horizon control approach used to generate artificial trajectories that preserve essential human characteristics while solving the tracking problem. Finally, results are presented in~\S~\ref{sec:results}, followed by conclusions and future research directions in~\S~\ref{sec:conclusions}.

\section{TickTacking Interface: Background and Human Features Extraction}
\label{sec:ticktacking}

\begin{figure}
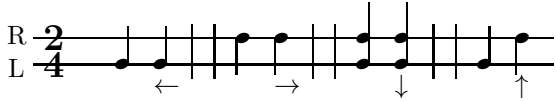

        \begin{music}
            \setsongraise1{4mm}
            \nobarnumbers
            \hsize=82mm
            \setclefsymbol{1}{\empty}%
            \nostartrule
            \generalmeter{\ccharnote {0}{L~~~~~}\ccharnote {5}{R~~~~~}\meterfrac{2}4 }
            \setlines{1}{2}
            \startextract%
            \Notes\smallnotesize\multnoteskip{0.8}\qu{2}\zsong{$\leftarrow$}\qu{2}\en\bar\bar  
            \Notes\smallnotesize\multnoteskip{0.8}\qa{6}\zsong{$\rightarrow$}\qa{6}\en\bar\bar
            \Notes\smallnotesize\multnoteskip{0.8}\zq{2}\qu{6}\zq{2}\zsong{$\downarrow$}\qu{6}\en\bar\bar  
            \Notes\smallnotesize\multnoteskip{0.8}\qu{2}\zsong{$\uparrow$}\qa{6}\en
            \zendextract
        \end{music}
        \caption{Encoding of controlled velocity directions, in case duplets are used by the TickTacking interface.}
        \label{fig:TickTacking}
        \vspace{-1.5em}
    \end{figure}

    TickTacking investigates rhythmic human control behavior through an interface consisting of a visual display and a dual-button input device, with each button assigned to one hand. Users control the direction and magnitude of a pointer's velocity on the interface screen by varying the combination and rate of their taps. In~\cite{rocchesso2024spacetime}, the use of \textit{duplets} and \textit{triplets} -- i.e., rhythmic patterns consisting of two or three taps per button within the same rhythmic cell -- was examined. When no tapping occurs, the pointer maintains a constant velocity. The interface design constrains movement by allowing only one velocity coordinate to change at a time. As an example, Fig.~\ref{fig:TickTacking} illustrates, using music notation, how different combinations of Right (R) and Left (L) button taps determine the controlled component (and its sign) of the pointer's velocity vector when encoded as duplets. In this encoding, the Inter-onset Interval (IoI) -- the time between two successive taps within the same duplet -- modulates the magnitude of the controlled velocity component at the end of each duplet. Specifically, the absolute value of the controlled velocity component is inversely proportional to the IoI. 

    In~\cite{rocchesso2024spacetime}, the effectiveness of the TickTacking interface was evaluated in two experiments with ten and seventeen participants, respectively. These experiments had different objectives: the first assessed controllability and user engagement, while the second examined interface usage under partial visual feedback deprivation. Given the additional complexity of visual feedback deprivation and the challenge of modeling an artificial controller to account for it, the present study focuses on the results of the first experiment from~\cite{rocchesso2024spacetime}. In the first experiment, ten participants first completed a training phase where they freely explored the TickTacking interface to navigate a two-dimensional space. They then performed a target-tracking task. Each participant used both versions of the interface: one utilizing duplets and the other triplets. Due to incomplete recordings of free navigation sessions, the final dataset consists of 16 free navigation trajectories and 20 tracking trajectories. Analysis of this data revealed the following key insights:
    \subsubsection{Velocity Direction} 
\begin{figure}
        \centering
        \vspace{0.5em}
        \includegraphics[width=0.72\linewidth]{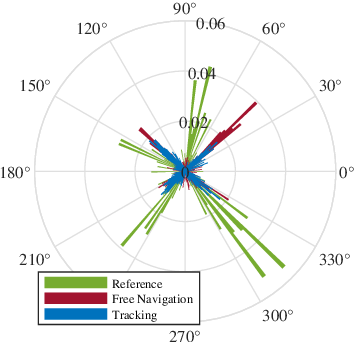}
        \caption{Histogram of velocity directions for the free navigation and target tracking scenarios in~\cite{rocchesso2024spacetime}, compared to the velocity directions of the reference trajectory. The bar height represents the relative frequency of observations.}
        \label{fig:human_total_directions}
        \vspace{-2em}
    \end{figure}

As pointed out in~\cite{rocchesso2024spacetime}, the histogram of velocity directions reveals that users predominantly chose specific directions, likely due to the ease of generating certain movements. Specifically, Fig.~\ref{fig:human_total_directions} shows that users favored directions at $45^\circ$, $135^\circ$, $225^\circ$ or $315^\circ$ relative to the positive horizontal axis in both free navigation and target tracking tasks. This suggests a preference for equal-length tapping time intervals along the horizontal and vertical axes. This preference aligns with findings in related fields such as animal locomotion~\cite{laffi2025rhythm}, vocal communication~\cite{de2021categorical}, and music cognition~\cite{jacoby2024commonality}, where simple temporal ratios (e.g., $1:1$, $1:2$, $2:1$) are commonly favored. These results indicate that human motor behavior in rhythmic control tasks is influenced by fundamental timing principles observed across different domains.
    \subsubsection{Velocity Magnitude} Another key human-related feature, already considered in~\cite{rocchesso2024spacetime}, is velocity magnitude, measured in pixels per second and originally defined as the Euclidean norm of the velocity vector. However, in this work, we instead define the velocity magnitude using its $l_1$ norm, which better aligns with the TickTacking interface, as velocity changes occur coordinate-wise. Additionally, the $l_1$ norm has the favorable numerical characteristic of being linear in the velocity components. Fig.~\ref{fig:human_total_velocities} presents a \gls{kde} of the velocity $l_1$ norm for both the free navigation and target tracking scenarios. 
    For analysis purposes, we approximate the free navigation~\gls{kde} with its best-fit Gaussian \gls{pdf}, itself illustrated in Fig~\ref{fig:human_total_velocities} with a dash-dot line. This Gaussian~\gls{pdf} closely aligns with the free navigation~\gls{kde} while providing a more convenient formulation. We observe that the~\gls{kde} of the human tracking data is a trade-off between the reference~\gls{kde} and the Gaussian \gls{pdf}, balancing the goal of perfect tracking with inherent human limitations. This suggests that humans strive to adapt their natural tapping tendencies~\cite{engler2024spontaneous} to achieve acceptable tracking performance.
    \subsubsection{Control Sparsity} 
\begin{figure*}
        \centering
        \vspace{0.5em}
        \subfigure[KDEs of the velocity $l_1$ norm in the free navigation and target tracking scenarios, compared to the KDE of the reference trajectory. The Gaussian PDF that best fits the free navigation data is also shown for comparison.]
        {
            \includegraphics[width=0.89\linewidth]{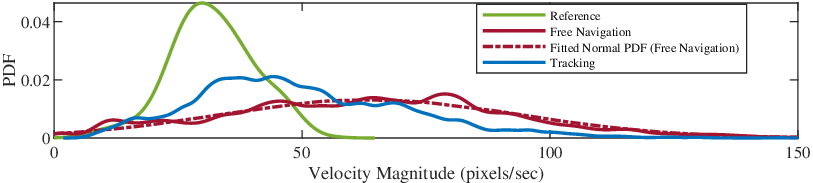}
            \label{fig:human_total_velocities}
        }
        \\
        \subfigure[Empirical CDF of the time intervals between control action updates in the free navigation and target tracking scenarios.]
        {
            \includegraphics[width=0.89\linewidth]{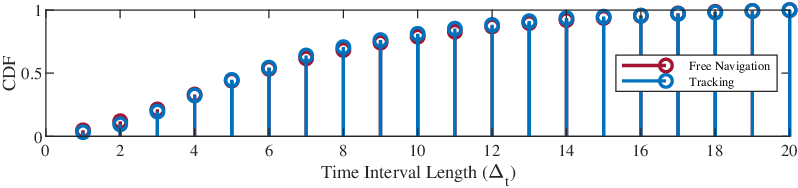}
            \label{fig:human_cdf}
        }
        \caption{Features of the human-generated trajectories acquired in~\cite{rocchesso2024spacetime} (in addition to those shown in Fig.~\ref{fig:human_total_directions}).}
        \label{fig:velocity_features}
        \vspace{-1.5em}
    \end{figure*}

In \cite{rocchesso2024spacetime}, human-generated trajectories using the TickTacking interface exhibited a characteristic zig-zag pattern. This pattern arises from the interface's design, its technical characteristics, and the dexterity required for effective operation, leading to both spatial and temporal sparsity in user input. We have already discussed the cause of spatial sparsity: the interface allows the user to affect only one velocity component at a time. Temporal sparsity, on the other hand, stems from the infrequent nature of user actions due to both biomechanical constraints and the time required by the users to observe the outcome of their previous action before choosing the next one. Fig.~\ref{fig:human_cdf} illustrates this feature through the empirical \gls{cdf} $\hat{F}_{T}$ of the time interval between input actions in the free navigation case. The time interval is measured in sample intervals $\Delta_t \in \mathbb{R}$, where $\Delta_t=0.283$ sec. The~\gls{cdf} is defined over the support $\{\Delta_t, \ldots, 20 \Delta_t\}$, with low-probability larger time intervals removed for refinement. Additionally, Fig.~\ref{fig:human_cdf} presents the corresponding empirical \gls{cdf} for the target tracking scenario, which closely aligns with the free navigation \gls{cdf}. This minimal divergence suggests that the interval length between control actions is a fundamental human property unaffected by the specific task or objective. 
    
    These findings define the ideal features for trajectories generated by a human-inspired artificial controller. As a result, these characteristics will inform the design of the optimization-based controller, guiding the formulation of its penalty terms and constraints.

\section{Human-Inspired Trajectory Generation Using Receding Horizon Tracking}
\label{sec:receding_horizon}
    The TickTacking tracking task can be seen as a control problem in which the user acts as a controller, seeking actions that best allow the pointer to follow the target. From this standpoint, a natural way to generate human-like artificial trajectories is to replace the user with a control algorithm designed to mimic human behavior as closely as possible. A suitable approach for this constrained control problem is the \gls{rh} controller, also known as a \gls{mpc}~\cite{rawlings2017model} algorithm. 
    
    The \gls{rh} technique is a recursive strategy. In a discrete-time framework, where $N$ is the final decision stage, and $W \in \{1, \ldots, N\}$ is the prediction window length (a key hyperparameter), the controller solves a constrained optimization problem at each time instant $t = n \Delta_t$, with $ n \in \{1, \ldots N - W + 1\}$, based on the current state $\mathbf{s}(t)$ of the dynamical system. This optimization determines the optimal control sequence $\mathbf{U}^{\star}(t) = \{\mathbf{u}^{\star}(t), \mathbf{u}^{\star}(t+\Delta_t), \ldots, \mathbf{u}^{\star}(t+(W-1)\Delta_t)\}$. Then, the first control action $\mathbf{u}^{\star}(t)$ is applied, updating the system state to $\mathbf{s}(t + \Delta_t)$,  the time window shifts forward by one unit (it ``recedes''), and the process repeats. 
    
    \gls{rh} controllers approximate the optimal solution of a \textit{global} optimization problem that would involve an infinite prediction window. However, solving such a problem is generally infeasible due to its infinite number of optimization variables. Indeed, balancing the length of the prediction window is a delicate act that must optimize the trade-off between achieving long-term objectives and guaranteeing computational tractability. Moreover, for the specific case addressed in this work, limiting the prediction horizon ensures that the algorithm does not gain an unrealistic advantage over human users, who typically predict future trajectory evolution within a finite time span.

    Since \gls{rh} controllers are \textit{model-based}, designing the controller first requires the definition of a kinematic model, specifying the state and action variables and their relationship through the state transition function. We consider a simple discrete-time state-space kinematic model to describe the motion of the pointer on the screen. The state $\mathbf{s} \in \mathbb{R}^{2}$ consists of the horizontal and vertical coordinates $x$ and $y$ of the pointer's position, while the control action $\mathbf{u} \in \mathbb{R}^{2}$ represents the velocity vector. The state transition function is
    \begin{equation}
        \label{eq:kinematic_system}
        \mathbf{s}(t + \Delta_t) = \mathbf{s}(t) +  \mathbf{u}(t)\,\Delta_t \,.
    \end{equation}
    For simplicity, we assume the sampling interval $\Delta_t$ corresponds to the time between any two successive possible control updates. The duration of $\Delta_t$ is assumed to be large enough to contain a single tap combination (duplets or triplets), if any, on the TickTacking interface, eliminating the need to explicitly distinguish between these rhythmic elements. 
    Additionally, since the TickTacking interface allows for no control update (when no button is pressed), we introduce an additional action variable $\mathbf{u}_{\textrm{old}} \in \mathbb{R}^{2}$ to retain the previous value of the controlled velocity. This variable is updated according to $\mathbf{u}_{\textrm{old}}(t+\Delta_t)=\mathbf{u}(t)$.
    
    Leveraging the variables that we have introduced, at each time step $t$, the~\gls{rh} controller solves the following mathematical programming problem:
    \begin{align}
    \label{eq:optimization_problem}
        &\min_{\mathbf{U}(t), \mathbf{S}(t), \mathbf{d}_s(t)}  \textrm{MSE}(\mathbf{S}(t), \mathbf{S}_{\textrm{ref}}(t)) + \sum_{k = 1}^{N_{p}} \lambda_{k}\mu_{k}(\mathbf{U}(t))\,, \\
        &\hspace{10pt} \textrm{s.t.} \hspace{1em}  \mathbf{g}\left(\mathbf{S}(t), \mathbf{U}(t), \mathbf{U}_{\textrm{old}}(t), \mathbf{d}_s(t)\right) \leq \mathbf{0}\,, \nonumber \\
        &\phantom{\hspace{10pt} \textrm{s.t.} \hspace{1em}} \mathbf{h}\left(\mathbf{S}(t), \mathbf{U}(t), \mathbf{U}_{\textrm{old}}(t), \mathbf{d}_s(t)\right) = \mathbf{0}\,. \nonumber
    \end{align}
    Here, $\mathbf{U}(t) = \{\mathbf{u}(t), \ldots, \mathbf{u}(t+(W-1)\Delta_t)\}$ represents the control sequence over the prediction window, $\mathbf{U}_{\textrm{old}}(t)$ is the corresponding sequence of $\mathbf{u}_{\textrm{old}}$, $\mathbf{S}(t) = \{\mathbf{s}(t), \ldots, \mathbf{s}(t+W\Delta_t)\}$ is the state sequence and $\mathbf{S}_{\textrm{ref}}(t)$ is the corresponding state sequence of the reference trajectory. The sequence $\mathbf{d}_s(t) = \{d_s(t), \ldots, d_s(t+(W-1)\Delta_t)\}$, with $d_s \in \mathbb{R}$, represents a sequence of additional decision variables to enforce specific controller features. 
    
    The objective function includes the~\gls{mse} term
    \begin{equation}
    \label{eq:mse}
        \textrm{MSE}\left(\mathbf{S}(t), \mathbf{S}_{\textrm{ref}}(t)\right) \triangleq \frac{1}{W}\sum_{k=1}^{W} \Vert \mathbf{s}(t + k \Delta_t) - \mathbf{s}_{\textrm{ref}}(t + k \Delta_t)\Vert^{2}_{2}\,,
    \end{equation}
    along with $N_{p}$ penalty terms $\mu_{k}$, each weighted by a non negative scalar $\lambda_{k}$, to enforce desired controller behaviors. The notation $\Vert \cdot \Vert_{2}$ denotes the Euclidean or $l_2$ norm.
    The constraints, represented by the vector-valued functions $\mathbf{g}(\cdot)$ and $\mathbf{h}(\cdot)$, ensure that the controller satisfies certain conditions.  Fundamental inequalities constraints bound the state components to the TickTacking interface dimensions, enforcing $0 \, \textrm{pixels} \leq x \leq 1920 \, \textrm{pixels}$ and $0 \, \textrm{pixels} \leq y \leq 1080 \, \textrm{pixels}$. 
    Among the key equality constraints are the kinematic system~\eqref{eq:kinematic_system} and the update rule for $\mathbf{u}_{\textrm{old}}$.

    \subsection{Human-Like RH Control}
    \label{subsec:con_pen_hlc}

    The optimal solution to the trivial \gls{rh} target tracking problem~\eqref{eq:optimization_problem}, constrained only to the kinematic system and the interface dimensions, and without penalty terms, produces a tracking trajectory that, after the first time step, aligns with the target trajectory, but deviates significantly from human motion. Consequently, this solution serves as a baseline for developing the human-inspired \gls{rh} controller by introducing specific human-related penalty terms, constraints, and modifications to the control strategy.
    
    Following the order in~\S~\ref{sec:ticktacking}, the first human-related feature we consider is the distribution of the velocity directions. Since the most common directions in human motion are $45^\circ$, $135^\circ$, $225^\circ$ and $315^\circ$ -- where velocity components have equal magnitude -- we introduce the penalty term: 
    \begin{equation*}
        \mu_{1}(\mathbf{U}(t)) = \frac{1}{W}\sum_{k=0}^{W-1} \Big\vert \vert v_{x}(t + k \Delta_t)\vert - \vert v_{y}(t + k \Delta_t)\vert \Big\vert.
    \end{equation*}
    We recall that $v_x$ and $v_y$ are the two components of the control $\mathbf{u}$. Then, the term $\mu_1$ penalizes deviations from the preferred diagonal directions, thereby encouraging human-like directional movement.

    Another important characteristic of human motion is the magnitude of velocity, represented by its $\ell_1$ norm. Based on the Gaussian described in \S~\ref{sec:ticktacking} and shown in Fig.~\ref{fig:human_total_velocities} -- with mean $63.75$ pixels/sec and standard deviation $30.45$ pixels/sec -- we define the penalty term $\mu_{2}$ as the following approximation of its negative log formulation:
    \begin{equation*}
        \mu_{2}(\mathbf{U}(t)) = \frac{1}{W}\sum_{k = 0}^{W-1} (\Vert \mathbf{u}(t+k\Delta_t)\Vert_{1}  - 63.75)^{2}.
    \end{equation*}
    This approximation is twofold: with respect to the negative log formulation of the Gaussian \gls{pdf}, we neglect the additional constant term $-\log(\sqrt{2\pi}\sigma_{\vert\mathbf{u}\vert})$ and drop the squared standard deviation at the denominator of $\mu_{2}$. By combining this term with $\mu_1$, we ensure that velocity magnitudes remain within human-like ranges.

    The last features to address concern the spatial and temporal sparsity of the human control, affecting the direction and frequency of the control updates, respectively. We begin by taking into account the spatial sparsity. 
    To model spatial sparsity, we introduce the integer decision variable $d_{s} \in {-1, 0, 1}$, enforcing the update of only one velocity component at a time via the following conditional constraints:
     \begin{equation*}
        \left\{
        \begin{array}{ll}
            d_{s} \leq 0.9 & \Rightarrow v_{x} = v_{x,\text{old}}\,, \\
            d_{s} \geq -0.9 & \Rightarrow v_{y} = v_{y,\text{old}}\,,
        \end{array}
        \right.
    \end{equation*}
    which bind the velocity update to the specific value of the corresponding $d_{s}$. These constraints ensure that velocity updates occur only along one axis at a time, also introducing a rudimentary temporal sparsity. To enforce temporal sparsity, we modify the control strategy to obtain a stochastic event-triggered formulation. Specifically, the optimization problem~\eqref{eq:optimization_problem} is solved and the control $\mathbf{u}$ updated only if the stochastic condition
    \begin{equation}
    \label{eq:stochastic_condition}
        d_{t} \leq \hat{F}_{T}(t_{\textrm{event}})\,
    \end{equation}
    is met. Here, $\hat{F}_{T}$ is the~\gls{cdf} described in \S~\ref{sec:ticktacking} and shown in Fig.~\ref{fig:human_cdf}. The variable $d_{t} \in [0, 1]$ is a uniformly distributed indicator variable, sampled at the first time instant and at any time instant in which the control action is actually updated. The variable $t_{\textrm{event}} \in \{\Delta_{t}, 2\Delta_{t}, \ldots, 20\Delta_{t}\}$ represents the time intervals since the last control action update, with its domain corresponding to the support of $\hat{F}_{T}$. Condition~\eqref{eq:stochastic_condition} ensures that the~\gls{rh} control strategy has an overall~\gls{cdf} of $t_\textrm{event}$ coincident with $\hat{F}_{T}$, thereby mimicking the temporal limitations associated with the human behavior. 

    Additionally, the length of the prediction window plays a crucial role in the controller’s temporal characterization, ensuring it does not outperform human users unfairly. We set the prediction window length to $8\Delta_{t}$, the integer closest to the mean of $\hat{F}_{T}$. In this way, the prediction window is, on average, equivalent to the time interval in which the control actions remain unchanged. This reasoning supports the introduction of the constraints
    \begin{equation*}
        \mathbf{u}(t) = \mathbf{u}(t + \Delta_t) = \ldots = \mathbf{u}(t + (W-1)\Delta_t)
    \end{equation*}
    which guarantee the identity between the elements of $\mathbf{U}(t)$.
    
    Finally, we observe that the formulation of the human-like~\gls{rh} controller, which incorporates all the key human features discussed in~\S~\ref{sec:ticktacking} through additional penalty terms and constraints, still maintains a simple theoretical interpretability. Moreover, from a computational perspective, the overall optimization problem remains quadratic. The use of the $\ell_1$ norm for velocity magnitude and the structured constraint design ensures that the problem remains efficiently solvable. This allows us to leverage specialized solvers such as \texttt{GUROBI} for the rapid optimization of~\eqref{eq:optimization_problem}.
    
\section{Numerical Results}
\label{sec:results}

    We analyze two scenarios: a baseline scenario, where we present the baseline \gls{rh} controller's results, and a human-like tracking scenario, where we examine the performance of the human-inspired \gls{rh} controller described in \S~\ref{subsec:con_pen_hlc}. To match the dataset size in~\cite{rocchesso2024spacetime}, we generate $20$ artificial trajectories in both scenarios initialized as each human-generated one. 
    
    The baseline scenario illustrates the behavior of the basic~\gls{rh} controller, which minimizes the~\gls{mse}~\eqref{eq:mse} at each time step, subject only to the kinematic system dynamics and display boundaries. In contrast, the second scenario explores the case where we enforce the human-inspired features described in~\S~\ref{sec:receding_horizon} to define the human-like~\gls{rh} controller. Table~\ref{tab:table} presents the manually tuned parameters associated with such controller: $\lambda_{1}$ and $\lambda_{2}$ are the weights of the penalty terms, and $\sigma_{\varepsilon}$ is the standard deviation of a Gaussian distributed, zero-mean noise added to the controlled action to account for residual stochasticity in human behavior not explicitly modeled by our formulation. The tuning process follows a trial-and-error approach aimed at replicating the observed statistical features of human behavior.
   \begin{table}
        \centering
        \vspace{1em}
        \renewcommand{\arraystretch}{1} 
        \setlength{\tabcolsep}{15pt} 
        \begin{tabular}{*{3}{c}}
        \toprule
        $\lambda_{1}$ & $\lambda_{2}$ & $\sigma_{\varepsilon}$ \\ 
        \midrule
        $32$    & $0.6$    & $2$ pixels/sec  \\ 
        \bottomrule
        \end{tabular}
        \caption{Tuned parameters of the Human-Like ~\gls{rh} Controller.}
        \label{tab:table}
        \vspace{-1.5em}
    \end{table}
    
\begin{figure}
       \centering
       \includegraphics[width=0.89\linewidth]{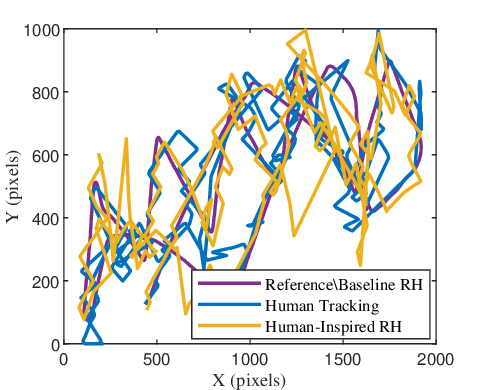}
       \caption{Example trajectories generated by the baseline and human-inspired RH controllers. The controller and reference trajectories practically overlap in the baseline scenario. One example of human-generated trajectories is included for comparison. A dynamic visualization of these trajectories is available at \href{https://doi.org/10.5281/zenodo.15240338}{https://doi.org/10.5281/zenodo.15240338}}
       \label{fig:examples_RH}
       \vspace{-1.5em}
   \end{figure}

\begin{figure}
       \centering
       \vspace{0.5em}
       \includegraphics[width=0.72\linewidth]{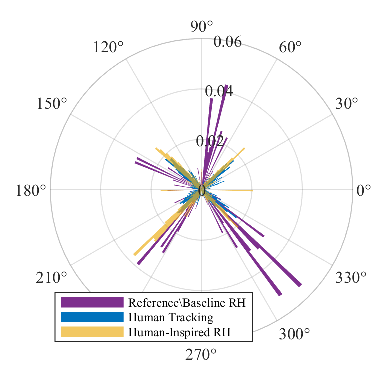}
       \caption{Histogram of velocity directions in the baseline~\gls{rh} and human-like~\gls{rh} scenarios, compared to the velocity directions in the human tracking scenario and the reference trajectory. The velocity directions of the reference trajectory and the baseline~\gls{rh} scenario overlap. Bars corresponding to the~\gls{rh} cases are partially transparent to enhance the visibility of the other two cases. The bar height represents the relative number of observations.}
       \label{fig:total_RH_directions}
       \vspace{-1.5em}
   \end{figure}

\begin{figure*}
        \centering
        \vspace{0.5em}
        \subfigure[KDEs of the velocity $l_{1}$ norm in the baseline~\gls{rh} and human-like~\gls{rh} scenarios. The KDEs of the human tracking scenario and the reference trajectory are shown for comparison.]
        {
            \includegraphics[width=0.89\linewidth]{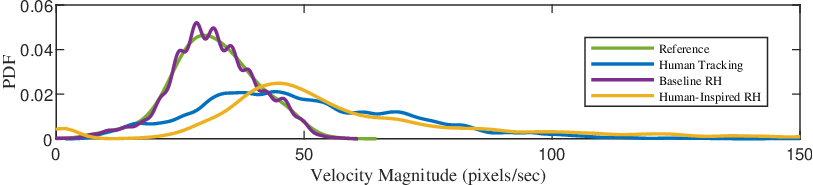}
            \label{fig:total_velocities}
        }
        \\
        \subfigure[KDEs of the MSE in the baseline~\gls{rh}, human-like~\gls{rh}, and human tracking scenarios.]
        {
            \includegraphics[width=0.89\linewidth]{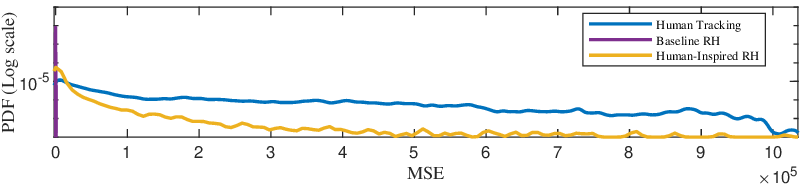}
            \label{fig:total_MSE}
        }
        \caption{Features of the~\gls{rh}-generates trajectories (in addition to those shown in Fig.~\ref{fig:total_RH_directions}).}
        \label{fig:results_features}
        \vspace{-1.5em}
    \end{figure*}

   The results of the baseline scenario are shown in Fig.~\ref{fig:examples_RH}, Fig.~\ref{fig:total_RH_directions} and Fig.~\ref{fig:results_features}. The velocity directions and $l_1$ norms of the baseline~\gls{rh} controller align closely with those of the reference trajectory. All generated trajectories quickly converge to the reference trajectory, achieving near-complete overlap as early as the second time step. The~\gls{kde} of the~\gls{mse} is impulsive, with the peak at zero. 
   
   The same figures also depict the results of the human-like scenario, showing the effect of the penalty terms $\mu_{1}$ and $\mu_{2}$. They facilitate the generation of velocities that closely resemble those observed in human-generated trajectories for the same tracking task. When combined with the sparse control strategy, they enable the controller to replicate the human behavior, deviating from the exact tracking exhibited by the baseline \gls{rh} controller. This affects coherently the~\gls{kde} of the~\gls{mse}, which still has a peak at zero but lacks the impulsive behavior seen in the baseline case. Nevertheless, the~\gls{kde} of the~\gls{mse} indicates that the human-like~\gls{rh} controller generally achieves lower~\gls{mse} values compared to the empirical~\gls{pdf} of human-generated trajectories.
   
   These findings suggest that the human-like~\gls{rh} controller is a viable approach for generating trajectories that resemble human behavior while demonstrating improved performance. As a result, this controller may serve as a valuable tool for designing new experiments to further investigate human behavior within the TickTacking interface, and to assist human users in decision-making and training applications.

\section{Conclusions and Further Extensions}
\label{sec:conclusions}
 
This study investigated the generation of human-like trajectories using an~\gls{rh} approach applied to the TickTacking interface, a novel rhythm-based HCI method. By analyzing user-generated trajectories, we identified key human-related features such as velocity directions, velocity magnitude~\gls{pdf}, and the probabilistic characterization of control sparsity. These features were used to design a controller that mimics human behavior, providing insights into the trade-offs, constraints, and objectives that humans navigate when interacting with this interface. While relying on the modeling of human-likeness introduced in this work for the specific TickTacking task, the control design procedure is generalizable to other tasks where specific human-like features can be defined and exploited. Our findings contribute to the broader goal of developing general-purpose rhythm-based human-machine interfaces. By understanding and replicating human behavior in control systems, it is possible to design more intuitive and effective user guidance systems that enhance overall interaction quality. 

Future research could explore additional human-related features and extend the current analysis, potentially involving methods like Granger's causality~\cite{gnecco2024measuring} to assess the effectiveness in performing the tracking task. Furthermore, data-driven approaches could reduce computational demands by allowing to learn the prediction over the~\gls{rh} control horizon~\cite{masti2020learning}, or by tuning a lightweight model reference~\gls{lqr} controller to approximate~\gls{rh} performance~\cite{masti2021tuning}. Additionally, techniques from mixed-integer optimization~\cite{zamponi2025certified} could be adapted to enhance the efficiency of the optimization step within the control problem.
Finally, a closed user-controller loop could be envisioned, wherein the controller adapts to the user while the user concurrently learns from the controller.

\bibliographystyle{IEEEtran}
\bibliography{references}

\end{document}